\documentclass[preprint,aps,nofootinbib]{revtex4}
\usepackage{}
\usepackage{graphicx}
\usepackage{amsmath}
\usepackage{amsfonts}
\usepackage{amssymb}
\usepackage{color}%
\usepackage{dcolumn}
\usepackage{indentfirst}
\providecommand{\U}[1]{\protect\rule{.1in}{.1in}}

\definecolor{lightgray}{rgb}{.7,.7,.7}

\definecolor{red}{rgb}{1,0,0}

\definecolor{blue}{rgb}{0,0,1}

\definecolor{purple}{rgb}{0.6,0.1,0.7}

\newcommand{\f}{\begin{equation}}
\newcommand{\ff}{\end{equation}}
\newcommand{\fa}{\begin{eqnarray}}
\newcommand{\ffa}{\end{eqnarray}}

\begin{document}
\title{Bouncing universe with modified dispersion relation}
\author{Wen-Jian Pan $^{1}$}
\email{wjpan_zhgkxy@163.com}
\author{Yong-Chang Huang $^{1}$}
\email{ychuang@bjut.edu.cn}
\affiliation{$^1$ Institute of theoretical physics,Beijing University of Technology,Beijing,100124,China}

\begin{abstract}
 In this paper, employing the modified dispersion relation,
 we have derived the general modified Friedmann equations
 and the corresponding modified entropy relations for the
 Friedmann-Robertson-Walker (FRW) Universe. In this setup,
 we find that when the big bounce happens, its energy scale
 and its corresponding modified entropy behavior
 are sensitive to the value of $k$. In contrast to
 the previous work with $k=0$, our work mainly demonstrates
 that the bouncing behavior for the closed Universe with $k=1$
 appears at the normal energy limit of the modified dispersion
 relation introduced, and when bouncing phenomenon is in presence,
 its modified entropy is just equal to zero. Surprisingly, when $k=-1$,
 the bouncing behavior is in absence.
\end{abstract}
\maketitle

\section{Introduction}
 General relativity plays an important role in deeply linking
 the geometric structure of spacetime with the distribution of matter sources.
 This remarkable theory provides a powerful perspective to understand
 the nature of gravity and describe the evolution of the Universe. With the cosmological principle,
 the standard cosmological model can be derived successfully
 from general relativity. However, the cosmological singularity problem
 cannot be avoided effectively in this model. To solve this singularity
 problem, theorists widely turn to appealing for the quantum gravity.
 Unfortunately, up to now, one has not yet established
 a complete and consistent quantum theory of gravity, which results in that
 the singularity problem cannot be solved at first principle level.
 As a result, people need to seek for some important quantum properties
 in string theory and loop quantum gravity as important mechanisms
 to relieve or solve this problem at a phenomenological level.

 Nowadays people generally believe that the combination of
 general relativity and quantum mechanics will provide a fundamental
 minimal length or the maximal energy, which is referred as
 a feature of quantum gravity. The developments of string theory
 and loop quantum gravity have greatly
 intensified this belief. In particular, based on Loop Quantum Gravity
 there have been great interests in some modification of dispersion
 relations \cite{Gambini:1998it,Smolin:2002sz,AmelinoCamelia:2003xp},
 and Generalized Uncertainty Principles(GUP)\cite{Maggiore:1993rv,Garay:1994en}
 have been taken into account in the literature on string theory
 \cite{Gross:1987ar,Amati:1988tn} and on noncommutative geometry\cite{Doplicher:1994zv}.
 The forms of the modified dispersion relation (MDR) and ones of GUP
 can be viewed as the alternative mechanisms to handle
 the black hole problems, such as the singularity, the ``thermodynamics",
 the evaporation and the ``information paradox" and so on.
 In the recent years, the implication and application of the effects
 of quantum gravity have attracted a great deal of attentions
 \cite{Coleman:1998ti,AmelinoCamelia:2000zs,AmelinoCamelia:2000mn,Magueijo:2002xx,Myers:2003fd,
 AmelinoCamelia:2004xx,AmelinoCamelia:2005ik,Ling:2005bp,
Ling:2005bq,Li:2008gs,Han:2008sy,Adler:2001vs,Custodio:2003jp,
Medved:2004yu,Bolen:2004sq,Nozari:2006ka,Myung:2006qr,
Park:2007az,Xiang:2009yq,Ali:2012mt,Amelino-Camelia:2013fxa,
Ali:2014hma,Chen:2014bva,Tawfik:2015kga} ( for recent reviews
we refer to \cite{Hossenfelder:2012jw,AmelinoCamelia:2008qg,
Tawfik:2014zca} ).

 In the semi-classical limit an intuitive scenario was proposed to
 avoid the cosmological singularity problem
 in standard cosmology by a big bounce \cite{Bojowald:2001xe,
 Ashtekar:2006rx} that was implemented by modifying the standard
Friedmann equation. Recently, this scenario has been
extensively investigated in various approaches \cite{Ling:2006az,Ling:2008sy,
Ling:2009wj,Bozza:2009jx,Cai:2009in,Battisti:2010he,Corichi:2011pg,
Garriga:2013cix,Battefeld:2014uga,Vakili:2014fga,Pedram:2015jsa}.
In particular, based on the assumption that the gravitational
theory can be described by entropy force \cite{Jacobson:1995ab},
which has been widely interested by many theorists
\cite{Ling:2009wj,Padmanabhan:2002sha,Kothawala:2007em,Cai:2005ra,
Akbar:2006mq,Cai:2008mh,Eling:2006aw,Sheykhi:2007gi,Ge:2007yu,
Gong:2007md,Wu:2007se,Guedens:2011dy,Sharif:2012zzd,Cai:2008ys,Ling:2013qoa},
the authors in \cite{Ling:2009wj} proposed
that an interesting setup was that the modified dispersion
relation prevented the energy density of
the matter contents from diverging at high energy level such that
the modified Friedmann equation with the bouncing effect
can be derived successfully from the Clausius relation,
where the singularity of the corresponding spacetime was free.

 In the previous work \cite{Ling:2009wj}, the authors had discussed
the relevant bouncing behavior and the corresponding entropy behavior
for spatial flat universe by the modified dispersion relation.
Now we wonder whether similar interesting behaviors for the cases
with non-vanishing $k$ will happen. We expect to provide an answer to
this problem in a parallel way. It turns out that using the modified
dispersion relation in \cite{Ling:2009wj} to prevent the energy
density of the matter from diverging, we can obtain
the general modified Friedmann equations and
the corresponding modified entropy relations for the FRW Universe.
When $k=0$, our results naturally reduce to that considered in \cite{Ling:2009wj}.
In particular, we find that the modified Friedmann equation
contains bouncing behavior and the corresponding entropy is
non-negative for the closed universe. More precisely,
its bounce appears at the normal energy bound of the MDR,
namely $E=\frac{\pi}{2\eta l_p}$. This implies that
when bouncing phenomenon is in presence, its entropy is just equal to zero,
which is quite different from the result in the previous work \cite{Ling:2009wj}.
In addition, we have noted that when $k=-1$, the bouncing behavior
is in absence. Our results have strongly suggested that $k$ plays an important role on
the bouncing behavior of the Universe.

Our paper is organized as follows: in Sect. II, we briefly
review on the description of entropy force
for the Friedmann-Robertson-Walker universe;
in Sect. III, our task is to determine the modified universe
equations with the MDR in detail and discuss the corresponding entropy
relations; in Sect. IV, we discuss the bouncing
effects of the universe with various values of $k$
in detail; Sect. V gives conclusions and discussions.
In Appendix, we present a detailed calculation to give out
the reason why the bouncing behavior for the closed universe occurs at $a_0=\eta l_p$.
Here we have set $\hbar=c=1$ for convenience.
\section{the entropy force description}
 In the section, we will briefly recall some ingredients for
deriving the Friedmann-Robertson-Walker universe from the relation of the entropy force.
Let us begin with introducing the Friedmann-Robertson-Walker metric which is
\begin{equation}
 ds^2=-dt^2+a^2(t)[\frac{dr^2}{1-kr^2}+r^2d\Omega^2],
\end{equation}
where $k$ values can be taken $-1, 0$ and $1$ which correspond to
 the open, flat and closed Universe, respectively, and $a(t)$
 is the scale factor of the Universe. This spacetime geometry implies that
 the apparent horizon \cite{Cai:2008ys} can be defined as
\begin{equation}
 r_A={1\over\sqrt{H^2+\frac{k}{a^2}}}.
\end{equation}
In analogy with black hole thermodynamics, for the apparent horizon we can correspondingly define its entropy as
\begin{equation}
S=\frac{A}{4G}=\frac{\pi}{G(H^2+\frac{k}{a^2})},
\end{equation}
as well as the temperature
\begin{equation}
T=\frac{1}{2\pi r_A},
\end{equation}
where $A=4\pi r^2_A$ is the apparent horizon area.
In the viewpoint of entropy force, utilizing the Clausius relation,
$\delta Q=T \delta S$
, and integrating the both sides of this relation, one can naturally obtain Friedmann
equation
\begin{equation}
H^2+\frac{k}{a^2}={8\pi G\over3}\rho_t,
\end{equation}
where $\rho_t$ is total energy density. Here we have successfully derived Friedmann equation from
the Clausius relation via assuming that our universe is  homogeneous and isotropic,
and treating one as a thermodynamical system in equilibrium.
It is obvious that the singularity still exists.
Thus, in the next sections, after taking a specific MDR into account,
we expect to apply the corresponding entropy force prescription to construct the
modified universe equations with any spatial curvature,
which may include the bouncing universe behavior, so that the singular problem can be free.
\section{modified universe equations with modified dispersion relation}
In this section, we mainly concentrate on the modified dispersion relation
how it causes the modification of the FRW Universe equations with
a spatial curvature $k$ in semi-classical limit.
Here we will introduce the quantum gravitational effect at
the phenomenological level. Let us start with the modified
dispersion relation. Based on Doubly Special Relativity,
a general modified dispersion relation proposed in \cite{AmelinoCamelia:2000mn,Magueijo:2002xx} has a following
form,
\begin{equation}\label{md}
E^2f^2(l_pE)-p^2g^2(l_pE)=m^2,
\end{equation}
where $E$, $p$ and $m$ are the energy, momentum and rest mass of the particle, respectively, $l^2_{p}=8\pi G$.
The above MDR has been widely used to study black hole physics
and rainbow universe \cite{AmelinoCamelia:2005ik,Ling:2005bp,
Ling:2005bq,Li:2008gs,Han:2008sy,Ling:2006az,Ling:2008sy}.
At low energy limit, namely $l_pE\ll1$, one naturally requires that
the two general functions $f(l_pE), g(l_pE)\sim 1$, so that the above dispersion relation reduces to
the standard Einstein energy-momentum relation. The authors of the previous work \cite{Ling:2009wj}, in fact,
 have chosen the two specific functions for the modified dispersion relation
\begin{equation}
f^2(l_pE)=\frac{\sin^2(\eta l_pE)}{(\eta l_pE)^2},\ \ \ \ \ \ g^2(l_pE)=1,
\end{equation}
then Eq.(\ref{md}) gives rise to the following form,
\begin{equation}
{1\over\eta l_{p}}\sin(\eta l_{p}E)=\sqrt{(p^2+m^2)},\label{mdr}
\end{equation}
where $\eta$ is dimensionless parameter which is bigger than zero.
They have omitted the negative energy branch. The right side
of the equation is not less than zero, so that the value
of energy is in the range $[0,\frac{\pi}{\eta l_{p}}]$. Obviously, this
MDR can be separated into two branches, namely $[0,\frac{\pi}{2\eta l_p}]$
and $[\frac{\pi}{2\eta l_p},\frac{\pi}{\eta l_p}]$, respectively. It is easy to find that
in the different branches, the monotonous behavior of their momentum
is remarkably distinguishing, as the energy is increasing.
More explicitly, for the particle energy in branch $[0,\frac{\pi}{2\eta l_p}]$,
its momentum $p$ is increasing as the energy is growing,
while for the particle energy in branch $[\frac{\pi}{2\eta l_p},\frac{\pi}{\eta l_p}]$,
 $p$ has a strangely and monotonously decaying behavior.
 Thus we call the latter branch as the anomalous energy region.
 In particular, for the rest mass $m=0$, in between the two branches,
 when particle energy passes through $E=\frac{\pi}{2\eta l_p}$,
 its momentum reaches a peak value, namely$1\over\eta l_p$; whereas
 when $E=\frac{\pi}{\eta l_p}$, its momentum $p$ vanishes.
In order to introduce the bouncing effect in the Universe,
we need to establish further the variation
relation of the above equation (\ref{mdr}) that has a following form
\begin{equation}
\delta E = \pm \frac{1}{\sqrt{1-\sin^2(\eta l_pE)}}\frac{p}{\sqrt{(p^2+m^2)}} \delta p,\label{vmdr}
\end{equation}
where the plus sign $``+"$ and the minus sign $``-"$ stand for choosing
the energy value in the range $[0,\frac{\pi}{2\eta l_p}]$,
and $[\frac{\pi}{2\eta l_p},\frac{\pi}{\eta l_p}]$, respectively.
For the high energy limit or massless particle,
the above equation can be expressed as
 \begin{equation}
\delta E = \pm \frac{1}{\sqrt{1-(\eta l_pp)^2}}\delta p\label{vmdr2}.
\end{equation}
Later, we will see that the above relation plays
an important role in deriving the modified universe equation
which may relieve the singularity problem.
When the modified dispersion relation is taken into account,
the modified Friedmann equation can be derived via an
integration relation in \cite{Cai:2008ys}
\begin{equation}
{8\pi G\over3}\rho=-\frac{\pi}{G}\int S'(A)(\frac{4G}{A})^2dA,\label{caur}
\end{equation}
where $S'(A)=\frac{dS}{dA}$ is related to the modified
dispersion relation, while for ordinary case it is a well-known constant,
namely $\frac{1}{4G}$. Now let us consider how such modified
ratio of the entropy to the area on the apparent cosmological
horizon is derived out, and then how it governs the modified
Friedmann equation such that the singularity of the universe
can be free. We treat our universe as a thermodynamical system in equilibrium
and suppose a quantum process in which a single
massless particle with energy $E$ escapes through its apparent
horizon. This causes correspondingly the minimal
entropy change \cite{AmelinoCamelia:2004xx,Ling:2005bq,Han:2008sy},
which should satisfy Clausius relation,
 \begin{equation}
\delta S_{min}=dS=\frac{\delta Q}{T}=\pm{2\pi\over\sqrt{1-\frac{4\pi(\eta l_p)^2}{A}}}\label{caur2},
\end{equation}
here we have used the identification relations $\delta Q \sim E \sim \delta E$
as well as $p\sim \delta p$, and the Heisenberg uncertainty relation
$\delta p \sim \frac{1}{\delta x} \sim \frac{1}{r_A}$ \cite{Adler:2001vs}. Moreover,
we have also admitted a physical fact that for the particle in the quantum process
there is an intrinsic uncertainty position as its Compton wavelength
that is identified with the apparent horizon of the universe.
In this quantum process, correspondingly, one assumes that causing the minimal area
change of the apparent horizon is $\delta A_{min}=l_p^2=8\pi G$ \cite{Ling:2009wj}(and references therein).
Identifying $dA$ with $\delta A_{min}$ and using Equation (\ref{caur2}), we have
 \begin{equation}\label{sprime}
S'(A)=\frac{dS}{dA}=\frac{\delta S_{min}}{\delta A_{min}}=\pm\frac{1}{4G}\frac{1}{\sqrt{1-\frac{4\pi(\eta l_p)^2}{A}}}.
\end{equation}
Let us first consider the plus case. Putting this equation into
the integration relation (\ref{caur}), we can obtain
the modified Friedmann equation,
\begin{equation}\label{kfr1}
{8\pi G\over3}\rho=-\frac{2}{\eta^2 l_p^2}\sqrt{1-\frac{4\pi\eta^2 l_p^2}{A}}+C_1,
\end{equation}
when $4\pi\eta^2 l_p^2 \ll A$, Eq.(\ref{kfr1}) should reduce to the standard Universe
equation, which is
${8\pi G\over3}\rho+\frac{\Lambda}{3}=H^2+\frac{k}{a^2},$
 thus the integral constant can be figured out
\begin{equation}
C_1=\frac{2}{\eta^2 l_p^2}-\frac{\Lambda}{3}.
\end{equation}
So the modified kinematic Friedmann equation can be determined as
 \begin{equation}\label{fme1}
{8\pi G\over3}\rho_t={8\pi G\over3}\rho+\frac{\Lambda}{3}=\frac{2}{\eta^2 l_p^2}(1-\sqrt{1-\frac{4\pi\eta^2 l_p^2}{A}}).
\end{equation}
 Making both sides of Eq.(\ref{fme1}) square and simplifying it, we can obtain an alternative form as
 \begin{equation}
H^2+\frac{k}{a^2}={8\pi G\over3}\rho_t(1-\frac{\rho_t}{\rho_c}),\label{fride1}
\end{equation}
where $\rho_c=\frac{12}{\eta^2l^4_p}$. With the use of the continuity
equation $\dot{\rho_t}+3H(\rho_t+P_t)=0$, the modified dynamical Friedmann equation can be calculated out
 \begin{equation}
\frac{\ddot{a}}{a}={8\pi G\over3}\rho_t(1-\frac{\rho_t}{\rho_c})-4\pi G(\rho_t+P_t)(1-2\frac{\rho_t}{\rho_c}),\label{dyfride}
\end{equation}
where $P_t$ is the total pressure.

The modified entropy-area relation for energy range $[0,\frac{\pi}{2\eta l_p}]$ can be
expressed as
 \begin{equation}\label{s1}
S_M=\frac{A}{4G}\sqrt{1-\frac{4\pi\eta^2l^2_p}{A}}+\frac{\pi\eta^2l^2_p}{G}\ln[\sqrt{\frac{A}{4\pi\eta^2l^2_p}}+\sqrt{\frac{A}{4\pi\eta^2l^2_p}-1}]+D_1,
\end{equation}
when $\eta\rightarrow0$, the modified entropy-area relation reduces
to the normal result, namely $\frac{A}{4G}$, which can give rise to
the integral constant $D_1=0$.

Similarly, for the energy range $[\frac{\pi}{2\eta l_p},\frac{\pi}{\eta l_p}]$ we can work out the modified Universe equation
\begin{equation}\label{kfr2}
{8\pi G\over3}\rho=\frac{2}{\eta^2 l_p^2}\sqrt{1-\frac{4\pi\eta^2 l_p^2}{A}}+C_2.
\end{equation}
Here we require that when $A=4\pi\eta^2l^2_p$ corresponding to the
energy value $\frac{\pi}{2\eta l_p}$, Eq.(\ref{kfr2})
can match smoothly with Eq.(\ref{kfr1}) so that $C_2$ should be
equal to $C_1$, namely $C_2=C_1=\frac{2}{\eta^2 l_p^2}-\frac{\Lambda}{3}$.
In the parallel way, with the use of the above equation and continuity equation, it is not hard to reproduce
the modified universe equations which are the same as Eqs.(\ref{fride1}) and (\ref{dyfride}), respectively.
Similarly, the corrected entropy relation also becomes
\begin{equation}\label{s2}
S_M=-\frac{A}{4G}\sqrt{1-\frac{4\pi\eta^2l^2_p}{A}}-\frac{\pi\eta^2l^2_p}{G}\ln[\sqrt{\frac{A}{4\pi\eta^2l^2_p}}+\sqrt{\frac{A}{4\pi\eta^2l^2_p}-1}]+D_2.
\end{equation}
Here we also require that the modified entropy-area relations in both
Eqs.(\ref{s1}) and (\ref{s2}) are the same, when $A=4\pi\eta^2l^2_p$.
Thus we can obtain the integral constants satisfy a relation $D_2=D_1=0$.
It is easy to check that when $A=4\pi\eta^2l^2_p$ the value of entropy-area
vanishes, namely $S_M=0$.
\section{the bouncing behaviors in the modified universe equations}
In this section, we mainly explore the bouncing behavior for
the modified Friedmann's equations. In the bouncing universe
scenarios, the most appealing feature is no initial singularity
of the universe. In these bouncing paradigms, ones usually argue
that there should be a nonsingular connection between the two
distinguishing phases that are the contraction phase and the
expansion phase in order. More explicitly, the evolution of
the cosmological scale factor from the contraction phase with $\dot{a}<0$ to the expansion
phase with $\dot{a}>0$, goes through its minimal value with non-vanishing value referred
as a critical point. In such process, some effects (quantum effects)
that become the dominant ingredients would prevent the universe
from collapsing into a singularity and then drive our universe to
accelerate expansion. As a result, at the critical point
the cosmological scale reaches the minimal nonzero value($a_0>0$)
which has $H_0=0$ and $\ddot{a}_0>0$\cite{Ling:2009wj}.
Thus, we can reasonably view them as the bouncing conditions.
After taking the first bouncing condition into account, namely $H_0=0$,
solving reversely modified Friedmann equation (\ref{fride1}) gives rise to
\begin{equation}
\rho_{t\pm}=\frac{1}{2}\rho_c(1\pm\sqrt{1-\frac{k\eta^2l^2_p}{a^2_0}}),\label{den}
\end{equation}
which should meet the constraints $\rho_{t\pm} > 0$ and $r_{A}\geq0$.
Here we have denoted the corresponding scale factor as $a_0$ when the Hubble parameter vanishes.
Making use of Eqs.(\ref{fride1}) and (\ref{den}), the modified dynamical
Friedmann equation is correspondingly represented by
\begin{equation}
\frac{\ddot{a}_0}{a_0}=\frac{k}{a^2_0}\pm4\pi G\sqrt{1-\frac{k\eta^2l^2_p}{a^2_0}}(\rho_{t\pm}+P_t),\label{dyfride2}
\end{equation}
where the signs $``\pm"$ in the above equation correspond to $\rho_{t\pm}$, respectively.
 There are two points what we would like to emphasize. Firstly, $\rho_{t-}$
is an ordinary energy density in the energy scale region ${[0,\frac{\pi}{2\eta l_p}]}$,
while $\rho_{t+}$ is anomalous energy density in $[\frac{\pi}{2\eta l_p},\frac{\pi}{\eta l_p}]$.
When taking energy density relation in (\ref{den}) to calculate the cosmic acceleration in between
the contracting phase and the expansion phase,
we have to use the corresponding relation in (\ref{dyfride2}). Secondly,
when $k\neq0$, this means there is an undetermined degree of the freedom, namely the scale factor $a_0$.
Now we are going to divide them into three cases,
namely $k=-1$, $k=0$ and $k=1$, respectively, and check
whether they really satisfy the constraints and the other bouncing condition or not.
\begin{itemize}
\item For $k=-1$, when requiring this case to satisfy the first bouncing condition, one easily finds that
$r_A$ becomes an imaginary number, which is not consistent with the constraint $r_A\geq0$.
Thus for this case the bouncing solution does not exist!

\item For $k=0$, we have
\begin{eqnarray}
\rho_{t+}=\rho_c ; \rho_{t-}=0.
\end{eqnarray}
When our Universe reaches high energy bouncing critical point
which corresponds to the total energy density is rather $\rho_{t+}=\rho_c$
 than $\rho_{t-}=0$,
 this gives out the dynamical Universe
equation which is
\begin{equation}
\frac{\ddot{a}_0}{a_0}=4\pi G(\rho_c+P_t)>0,
\end{equation}
it is just the result of the previous work\cite{Ling:2009wj}.
\item For $k=1$ the total energy density equation (\ref{den}) naturally reduces to
the following form
\begin{equation}\label{root}
\rho_{t\pm}=\frac{\rho_c}{2}(1\pm \sqrt{1-\frac{\eta^2l^2_p}{a^2_0}}).
\end{equation}
It seems  that there are two bouncing solutions.
Based on the definition of the apparent horizon and the MDR, we can find
the scale factor is no less than $\eta l_p$.
In Appendix, according to the bounce conditions, we can show that
the bouncing effect of the Universe is presented at the minimum
scale value that means the momentum reaches the maximum value
corresponding to the critical point $E=\frac{\pi}{2\eta l_p}$
such that in fact the above equation (\ref{root}) can only take the same value,
namely $\rho_{t-}=\rho_{t+}=\frac{\rho_c}{2}$ with $a^2_0=\eta^2l^2_p$.
After the big bounce of our Universe happening, the total density
will decrease monotonously with our Universe expanding.
When the total energy density climbs up the value
$\rho_{t-}=\rho_{t+}=\frac{\rho_c}{2}$, Eq.(\ref{dyfride2}) can be rewritten as
\begin{equation}\label{aa}
\frac{\ddot{a}_0}{a_0}=\frac{1}{\eta^2l^2_p}>0.
\end{equation}
Thus for this case the bounce of the Universe can exist
in the normal energy region $[0,\frac{\pi}{2\eta l_p}]$ without
anomalous negative entropy!
\end{itemize}
Here it is worth noting that for the case $k=0$ the bouncing solution
 occurs at the critical point with maximum energy value $\frac{\pi}{\eta l_p}$,
but with negative entropy in \cite{Ling:2009wj}. For $k=1$
the bouncing phenomenon can be presented at the energy value
$\frac{\pi}{2\eta l_p}$ corresponding to the maximum momentum $1\over\eta l_p$
. This means that the modified entropy-area relation
is rather Eq.(\ref{s1}) than Eq.(\ref{s2}).
More precisely, when the universe performs its bouncing
behavior in high energy limit, its corresponding entropy-area
relation is just equal to zero. After the big bounce, its modified entropy
will increase monotonously with our Universe expanding.
For $k=-1$ the bouncing one is absent, since
it can not meet the constraint condition $r_{A}\geq0$.
\section{ conclusions and discussions }
In this paper, along with the spirit in \cite{Ling:2009wj},
assuming that both the Clausius relation
and the modified dispersion relation keep effective on the
apparent horizon, we have derived the general modified universe equations
and the corresponding modified entropy relations for the FRW Universe.
From the modified universe equations,
we find that the bouncing behavior is presence
for the closed Universe, besides the spatial flat Universe,
so that the cosmological singularity problem can
be solved for these cases at the phenomenological level.
In particular, in contrast to the bouncing solution
of the spatial flat universe with $k=0$
to occur at the maximum energy value $\frac{\pi}{\eta l_p}$ of the MDR (\ref{mdr}),
but with negative entropy in \cite{Ling:2009wj}, we have shown that
the bouncing behavior of the closed Universe with $k=1$
explicitly appears at the energy value
$\frac{\pi}{2\eta l_p}$, without negative entropy.
More precisely, when the closed universe performs the bouncing
behavior at the critical energy point $\frac{\pi}{2\eta l_p}$,
the corresponding entropy-area relation is just equal to zero.
Surprisingly the bouncing solution of $k=-1$ is absent, since
it can not satisfy the constraint condition $r_{A}\geq0$.

 However, it is well-known that the entropy relation of
the Universe is an open question in the bouncing or cyclic
universe scenarios, if we respect the second law of the thermodynamics
saying that the total entropy of the universe never decreases
as time flows. When the closed Universe evolves from
the contraction phase to the expansion phase, Eq.(\ref{s1})
suggests that the corresponding entropy-area relation
has diminishing behavior and increasing behavior, respectively.
These facts tell us that the entropy behavior in
the contraction phase does not agree with the second law,
but one in the expansion phase agrees with the second law. Thus,
in analogy with the previous bouncing or cyclic universe scenarios,
the well-known entropy problem that violates the second law
is still in presence in our scenario.
A deep understanding on how to link the diminishing
behavior of the entropy in the contracting phase and the second law
of the thermodynamics is still lacking.
There might be an approach to solve this problem via considering
the generalized second law of thermodynamics which has been discussed in many universe
models\cite{Izquierdo:2005ku,Setare:2008bb,Karami:2009yd,Sharif:2013bla}.
Under complying with the second law of the thermodynamics,
 it should be interesting to seek for some new modified dispersion
relations and new physical mechanism
to overcome or solve the singularity problem of the Universe.
 This aspect is left for future works.

 \section*{Acknowledgments}

We are grateful to the anonymous referee for helpful suggestions.
The work is supported by National Natural Science Foundation of
China (No. 11275017 and No. 11173028).

\section*{Appendix}
In this Appendix, we present a detailed calculation which demonstrates that
the bouncing solution for the closed universe can appear at the minimum of scale factor.
To obtain the scale factor size at the critical bouncing point,
we need to take into account the behavior of the total energy density.
 When our universe evolves to around the
critical point in high energy limit in the contracting phase,
we are interested to consider the behavior of total energy density
with the constant state parameter $\omega$ that takes the values
in the range $(-\frac{1}{3}, \infty)$. Since in this case
the cosmological constant density is very small and can be totally ignored,
the behavior of the universe is governed by the dominated-content with the constant state parameter $\omega$.
Thus for any given the state parameter value $\omega$ in the allowed region, we can identify
the dominated-content density with the total energy density. From
the continuity equation, we can get the following density relation
\begin{equation}\label{rho}
\rho_t=\frac{C}{a^{3(1+\omega)}},
\end{equation}
where $C$ is the integral constant.
From the Friedmann equation(\ref{fride1}), it gives rise to
\begin{equation}
(\frac{da}{dt})^2=\frac{8\pi G}{3}a^2\rho_t(1-\frac{\rho_t}{\rho_c})-1 \geq 0.
\end{equation}
When the above equation is equal to $0$, it means the Hubble parameter $H$ is zero
which is just the first bouncing condition.
Substituting Eq.(\ref{rho}) into the above equation, we can obtain
\begin{equation}\label{fried3}
36a^{4+6\omega}-12l^2_pCa^{3+3\omega}+\eta^2l^6_pC^2 \leq 0,
\end{equation}
for $\rho_t<\rho_c$ and $a\geq\eta l_p$ this equation should be valid. Now
let us determine the integral constant $C$. When putting $a=\eta l_p$
into Eq.(\ref{fried3}), we can easily get
\begin{equation}\label{a01}
\eta^2l^6_p(C-6\eta^{1+3\omega}l^{3\omega-1}_p)^2 \leq 0,
\end{equation}
on the other hand, the above equation itself has
\begin{equation}\label{a02}
\eta^2l^6_p(C-6\eta^{1+3\omega}l^{3\omega-1}_p)^2 \geq 0,
\end{equation}
it is worth noting that this is the constraint exactly what we are looking for.
This relation plays an important role in deciding the critical point.
As a consequence, combining Eqs.(\ref{a01})and(\ref{a02}), we have
\begin{equation}\label{intc}
C=6\eta^{1+3\omega}l^{3\omega-1}_p.
\end{equation}
Plugging it into Eq.(\ref{fried3}), we find
\begin{equation}\label{frid1}
a^{4+6\omega}-2(\eta l_p)^{1+3\omega}a^{3+3\omega}+(\eta l_p)^{4+6\omega} \leq 0.
\end{equation}
Eqs.(\ref{a01})and(\ref{a02}) suggest that $a_0=\eta l_p$ is a root of
the Friedmann equation (\ref{fride1}) satisfying the first bouncing
condition. It is easy to check that this root saturates totally the
other bouncing condition, as showed in Eq.(\ref{aa}). However this story
is not yet end so far, since Eq.(\ref{frid1})generally has other roots.
Therefore, in the next step, we have to show that other roots can not satisfy
the second bouncing condition, otherwise the bouncing behavior occurs at the
other roots. Here we provide a general trick to discuss this problem
without specific state parameter $\omega$.
For convenience we define a function
\begin{equation}
f(a)= a^{4+6\omega}-2(\eta l_p)^{1+3\omega}a^{3+3\omega}+(\eta l_p)^{4+6\omega}.
\end{equation}
In order to obtain the monotonicity of this function, we need to work out
 the derivative of the function with respective to the scale factor $a$,
which is
\begin{equation}
f'\equiv\frac{df(a)}{da}= (4+6\omega)a^{2+3\omega}[a^{1+3\omega}-\frac{3+3\omega}{2+3\omega}(\eta l_p)^{1+3\omega}]=0,
\end{equation}
for $\omega>-\frac{1}{3}$ and $a>0$, we can obtain
\begin{equation}
\bar{a}=(\frac{3+3\omega}{2+3\omega})^{\frac{1}{1+3\omega}}\eta l_p.
\end{equation}
Thus, we can easily check that $f(a)$ decreases  monotonously,
when the scale factor $a$ in the range $[\eta l_p, \bar{a}]$;
$f(a)$ increases  monotonously, when the scale factor $a$ is outside $\bar{a}$.
Since $a_0=\eta l_p$ is a root of $f(a)=0$ and $f(\bar{a})<0$, if it has other roots, then they must
be greater than $\bar{a}$.

Now we turn to considering the role of the second bouncing condition.
In order to seek for the bouncing solution in normal energy region,
we should consider the minus cases in both Eqs. (\ref{den})and(\ref{dyfride2}).
Substituting $\rho_{t-}=\frac{1}{2}\rho_c(1-\sqrt{1-\frac{\eta^2l^2_p}{a^2_0}})$ into Eq.(\ref{dyfride2}), we
find
\begin{equation}
\frac{\ddot{a}_0}{a_0}=\frac{3(1+\omega)}{\eta^2l^2_p}-\frac{2+3\omega}{a^2_0}-\frac{3(1+\omega)}{\eta^2l^2_p}\sqrt{1-\frac{\eta^2l^2_p}{a^2_0}}
\end{equation}
Requiring that the above equation is greater than zero, we can derive an important constraint
\begin{equation}
a^2_0<{(3\omega+2)^2\eta^2l^2_p\over(3\omega+3)(3\omega+1)}.
\end{equation}
It tells us that if any one root $a_0$ meets the above requirement,
then the bouncing behavior can appear at the root.
It is easy to check that $a_0=\eta l_p$ satisfies this constraint.
In following we will prove that
the other roots do not satisfy the requirement.
For the  purpose, we can define a function which has a following form
\begin{eqnarray}
g(\omega)&=&\frac{{(3\omega+2)^2\eta^2l^2_p\over(3\omega+3)(3\omega+1)}}{\bar{a}^2}.\nonumber\\
         &=&\frac{(3\omega+2)^{\frac{4+6\omega}{3\omega+1}}}{(3\omega+3)^{\frac{3+3\omega}{3\omega+1}}(3\omega+1)},
\end{eqnarray}
then taking it logarithm, we have
\begin{eqnarray}\label{yo}
y(\omega)&=&(3\omega+1)\ln{g(\omega)}\nonumber\\
         &=&(3\omega+3)[\ln(3\omega+2)-\ln(3\omega+3)]+(3\omega+1)[\ln(3\omega+2)-\ln(3\omega+1)].
\end{eqnarray}
To determine the monotonous behavior of this function, we can find
its derivative with respective to the state parameter $\omega$.
\begin{equation}
\frac{dy(\omega)}{d\omega}=3\ln\frac{(3\omega+2)^2}{(3+3\omega)(1+3\omega)}>0.
\end{equation}
Thus the function $y(\omega)$ increases monotonously in the range $(-\frac{1}{3},\infty)$.
Now we work out the behavior of the function at the infinity. Making Eq.(\ref{yo})take limit, we
can have
\begin{eqnarray}
\lim_{\omega\rightarrow\infty}y(\omega)&=&\lim_{\omega\rightarrow\infty}[(3\omega+3)(\ln\frac{3\omega+2}{3\omega+3})+
(3\omega+1)(\ln\frac{3\omega+2}{3\omega+1})]\nonumber\\
                                       &=&\lim_{\omega\rightarrow\infty}[-\frac{1}{6+6\omega}-\frac{1}{2+6\omega}+\ldots]\nonumber\\
                                       &=&0.
\end{eqnarray}
So, the function $y(\omega)$ is less than zero in $(-\frac{1}{3},\infty)$, which means that the
function $g(\omega)$ is less than $1$. This tells us that although the other roots can
exist, they can not saturate the second bouncing condition, namely $\ddot{a}_0>0$.
Thus, the bouncing solution appears at $a_0=\eta l_p$.

\end{document}